\documentclass[conference, a4paper]{IEEEtran}

\usepackage{algorithm,algorithmic}
\usepackage{amsmath,amssymb,mathrsfs}
\usepackage{array}
\usepackage{balance}
\usepackage{booktabs}
\usepackage{cite}
\usepackage{fancyhdr}
\usepackage{float}
\usepackage[nomain,acronym,automake,toc,nopostdot]{glossaries}%[nonumberlist,acronym,toc]
\usepackage{graphicx,color}
\usepackage{mathtools}
\usepackage{multirow}
\usepackage{psfrag}
\usepackage{stfloats}
\usepackage{subfigure}

\newtheorem{thm}{Theorem}
\newtheorem{cor}{Corollary}[thm]

\newtheorem{rmk}{Remark}

\makeglossaries
\newacronym[description=Additive white Gaussian noise]{awgn}{AWGN}{additive white Gaussian noise}
\newacronym[description= Alternating direction method of multipliers]{admm}{ADMM}{ alternating direction method of multipliers}
\newacronym[description=Approximate message passing]{amp}{AMP}{approximate message passing}
\newacronym[description={\em a posteriori} probability]{app}{APP}{{\em a posteriori} probability}
\newacronym[description= Alternative unbiased-preconditioning for iterative mMIMO
detection]{aupid}{AUPID}{alternative unbiased-preconditioning for iterative mMIMO
detection}
\newacronym[description=Base station]{bs}{BS}{base station}
\newacronym[description=Base station sleeping ]{bss}{BSS}{base station sleeping }
\newacronym[description=Belief propagation]{bp}{BP}{belief propagation}
\newacronym[description=Broyden-Fletcher-Goldfarb-Shanno]{bfgs}{BFGS}{Broyden-Fletcher-Goldfarb-Shanno}
\newacronym[description=Binary phase shift keying]{bpsk}{BPSK}{binary phase shift keying}
\newacronym[description=Bit error rate]{ber}{BER}{bit-error-rate}
\newacronym[description=Block error rate]{bler}{BLER}{block error rate}
\newacronym[description=Central limit theorem]{clt}{CLT}{central limit theorem}
\newacronym[description=Channel state information ]{csi}{CSI}{channel state information }
\newacronym[description=Closest vector problem]{cvp}{CVP}{closest vector problem}
\newacronym[description=Code division multiple access]{cdma}{CDMA}{code division multiple access}
\newacronym[description=Distributed linear data fusion]{dldf}{DLDF}{distributed linear data fusion}
\newacronym[description=European Cooperation in Science and Technology]{cost}{COST}{Cooperation in Science and Technology}
\newacronym[description=Coordinated multi-point ]{CoMP}{CoMP}{coordinated multi-point }
\newacronym[description=Correlation-based stochastic model]{cbsm}{CBSM}{correlation-based stochastic model}
\newacronym[description=Cumulative distribution function]{cdf}{CDF}{cumulative distribution function}
\newacronym[description=Degrees of freedom]{dof}{DoF}{degrees of freedom}
\newacronym[description=Element-based lattice reduction]{elr}{ELR}{element-based lattice reduction}
\newacronym[description=Extremely-large aperture array]{elaa}{ELAA}{extremely-large aperture array}
\newacronym[description=Fifth-generation]{5g}{5G}{fifth-generation}
\newacronym[description=Fixed-complexity sphere decoder]{fcsd}{FCSD}{fixed-complexity sphere decoder}
\newacronym[description=Forward error corrrection]{fec}{FEC}{forward error correction}
\newacronym[description=Free space path loss]{fspl}{FSPL}{free space path loss}
\newacronym[description=Gauss-Seidel]{gs}{GS}{Gauss-Seidel}
\newacronym[description=Global system for mobile communication]{gsm}{GSM}{global system for mobile communication}
\newacronym[description=Geometry-based stochastic model]{gbsm}{GBSM}{geometry-based stochastic model}
\newacronym[description=Gradient descent]{gd}{GD}{gradient descent}
\newacronym[description=Hermite-Korkin-Zolotarev]{hkz}{HKZ}{Hermite-Korkin-Zolotarev}
\newacronym[description=Independent and identically distributed]{iid}{i.i.d.}{independent and identically distributed}
\newacronym[description=Independent and non-identically distributed]{ind}{i.n.d.}{independent and non-identically distributed}
\newacronym[description=Integer least-squares]{ils}{ILS}{integer least-squares}
\newacronym[description=Iterative discrete estimation]{ide}{IDE}{iterative discrete estimation}
\newacronym[description=Iterative discrete estimation 2]{ide2}{IDE2}{iterative discrete estimation 2}
\newacronym[description=International Telecommunication Union Radiocommunication Sector ]{itu-r}{ITU-R}{International Telecommunication Union Radiocommunication Sector}
\newacronym[description=Large system behaviour]{lsb}{LSB}{large system behaviour}
\newacronym[description=Lattice reduction]{lr}{LR}{lattice reduction}
\newacronym[description=Lenstra-Lenstra-Lov\'{a}sz]{lll}{LLL}{Lenstra-Lenstra-Lov\'{a}sz}
\newacronym[description=Likelihood ascent search]{las}{LAS}{likelihood ascent search}
\newacronym[description=Line-of-slight]{los}{LoS}{line-of-slight}
\newacronym[description=List sphere decoder]{lsd}{LSD}{list sphere decoder}
\newacronym[description=Linear minimum mean square error]{lmmse}{LMMSE}{linear minimum mean square error}
\newacronym[description=Log-likelihood ratio]{llr}{LLR}{log-likelihood ratio}
\newacronym[description=Long-term evolution ]{lte}{LTE}{long-term evolution}
\newacronym[description=Low density parity check]{ldpc}{LDPC}{low density parity check}
\newacronym[description=Massive machine type communications]{mmtc}{mMTC}{massive machine type communications}
\newacronym[description=Maximum {\em a posteriori}]{map}{MAP}{maximum {\em a posteriori}}
\newacronym[description=Maximum likelihood]{ml}{ML}{maximum likelihood}
\newacronym[description=Maximum likelihood sequence detection]{mlsd}{MLSD}{maximum likelihood sequence detection}
\newacronym[description=Maximum ratio combining]{mrc}{MRC}{maximum ratio combining}
\newacronym[description=Multiple-input multiple-output]{mimo}{MIMO}{multiple-input multiple-output}
\newacronym[description=Massive multiple-input multiple-output]{mmimo}{mMIMO}{massive multiple-input multiple-output}
\newacronym[description=Matched filter]{mf}{MF}{matched filter}
\newacronym[description=Mean square error]{mse}{MSE}{mean square error}
\newacronym[description=Minimum mean square error]{mmse}{MMSE}{minimum mean square error}
\newacronym[description=Mobile and wireless communications Enablers for the Twenty-twenty Information ]{metis}{METIS}{Mobile and wireless communications Enablers for the Twenty-twenty Information}
\newacronym[description=Multi-user]{mu}{MU}{multi-user}
\newacronym[description=Non-line-of-sight]{nlos}{NLoS}{non-LoS}
\newacronym[description=Non-deterministic polynomial-time hard]{nphard}{NP-hard}{non-deterministic polynomial-time hard}
\newacronym[description=One dimensional]{1d}{1-D}{one dimensional}
\newacronym[description=Orthogonality defect]{od}{OD}{orthogonality defect}
\newacronym[description=Pairwise error probability]{pep}{PEP}{pairwise error probability}
\newacronym[description=Parallel interference cancellation]{pic}{PIC}{parallel interference cancellation}
\newacronym[description=Preconditioning for iterative mMIMO detection]{pid}{PID}{preconditioning for iterative mMIMO detection}
\newacronym[description=Probabilistic data association]{pda}{PDA}{probabilistic data association}
\newacronym[description=Probability distribution function]{pdf}{PDF}{probability density function}
\newacronym[description=Probability mass function]{pmf}{PMF}{probability mass function}
\newacronym[description=Quadrature amplitude modulation]{qam}{QAM}{quadrature amplitude modulation}
\newacronym[description=Quadrature phase shift keying]{qpsk}{QPSK}{quadrature phase shift keying}
\newacronym[description=Receiver-side channel state information]{rcsi}{R-CSI}{receiver-side channel state information}
\newacronym[description=Received signal strength]{rss}{RSS}{received signal strength}
\newacronym[description=Semidefinite relaxation]{sdr}{SDR}{semidefinite relaxation}
\newacronym[description=Seysen's algorithm]{sa}{SA}{Seysen's algorithm}
\newacronym[description=Signal-to-interference-plus-noise ratio]{sinr}{SINR}{signal-to-interference-plus-noise ratio}
\newacronym[description=Signal to interference ratio]{sir}{SIR}{signal to interference ratio }
\newacronym[description=Signal-to-noise ratio]{snr}{SNR}{signal-to-noise ratio}
\newacronym[description=Single antenna interference cancellation]{saic}{SAIC}{single antenna interference cancellation}
\newacronym[description=Single input single output]{siso}{SISO}{single input single output}
\newacronym[description=Singular value decomposition ]{svd}{SVD}{singular value decomposition }
\newacronym[description=Sixth-generation mobile networks]{6g}{6G}{sixth-generation}
\newacronym[description=Sphere decoder]{sd}{SD}{sphere decoder}
\newacronym[description=Space-time codes]{stc}{STC}{space-time codes}
\newacronym[description=State-of-the-art]{sota}{SoTA}{state-of-the-art}
\newacronym[description=Successive interference cancellation]{sic}{SIC}{succesive interference cancellation}
\newacronym[description=Symbol error rate]{ser}{SER}{symbol error rate}
\newacronym[description=Successive over-relaxation]{sor}{SOR}{successive over-relaxation}
\newacronym[description=Symmetric positive-definite]{spd}{SPD}{symmetric positive-definite}
\newacronym[description=Symmetric successive over-relaxation]{ssor}{SSOR}{symmetric successive over-relaxation}
\newacronym[description=Tabu search]{ts}{TS}{tabu search}
\newacronym[description=Three-dimensional]{3d}{3-D}{three-dimensional}
\newacronym[description=The 3rd Generation Partnership Project]{3gpp}{3GPP}{the 3rd Generation Partnership Project}
\newacronym[description=Two-dimensional]{2d}{2-D}{two-dimensional}
\newacronym[description=Uniform linear array]{ula}{ULA}{uniform linear array}
\newacronym[description=Urban micro]{umi}{UMi}{urban micro}
\newacronym[description=User equipment]{ue}{UE}{user equipment}
\newacronym[description=Vector error rate]{ver}{VER}{vector error rate}
\newacronym[description=Vertical Bell Labs layered space-time]{vblast}{V-BLAST}{vertical Bell Labs layered space-time}
\newacronym[description=Visibility region]{vr}{VR}{visibility region}
\newacronym[description=Widely linear]{wl}{WL}{widely linear}
\newacronym[description=Widely linear zero forcing]{wlzf}{WLZF}{widely linear zero forcing}
\newacronym[description=Wide-sense stationary uncorrelated scattering]{wssus}{WSSUS}{wide-sense stationary uncorrelated scattering}
\newacronym[description=Wireless World Initiative New Ratio]{winner}{WINNER}{Wireless World Initiative New Ratio}
\newacronym[description=Zero forcing]{zf}{ZF}{zero forcing}
\newacronym[description=Zero mean complex circularly symmetric]{zmccs}{ZMCCS}{zero mean complex circularly symmetric}

\definecolor{sblue}{RGB}{0,51,120}
\definecolor{sred}{RGB}{200,51,130}

\newcommand{\mat}[1]{\mbox{\boldmath $#1$}}

\newcommand{\figref}[1]{Fig. \ref{#1}}

\newcommand{\secref}[1]{Section \ref{#1}}
\renewcommand{\eqref}[1]{(\ref{#1})}

\ifCLASSINFOpdf
\else
\fi

\begin{document}
\title{Leveraging User-Wise SVD for Accelerated Convergence in Iterative ELAA-MIMO Detections}
\author{Jiuyu Liu, Yi Ma$^\dag$, and Rahim Tafazolli\\
	{\small 5GIC and 6GIC, Institute for Communication Systems, University of Surrey, Guildford, UK, GU2 7XH}\\
	{\small Emails: (jiuyu.liu, y.ma, r.tafazolli)@surrey.ac.uk}}
\markboth{}%
{}

\maketitle

\begin{abstract}
Numerous low-complexity iterative algorithms have been proposed to offer the performance of linear multiple-input multiple-output (MIMO) detectors bypassing the channel matrix inverse.
These algorithms exhibit fast convergence in well-conditioned MIMO channels. 
However, in the emerging MIMO paradigm utilizing extremely large aperture arrays (ELAA), the wireless channel may become ill-conditioned because of spatial non-stationarity, which results in a considerably slower convergence rate for these algorithms.
In this paper, we propose a novel ELAA-MIMO detection scheme that leverages user-wise singular value decomposition (UW-SVD) to accelerate the convergence of these iterative algorithms. 
By applying UW-SVD, the MIMO signal model can be converted into an equivalent form featuring a better-conditioned transfer function.
Then, existing iterative algorithms can be utilized to recover the transmitted signal from the converted signal model with accelerated convergence towards zero-forcing performance.
Our simulation results indicate that proposed UW-SVD scheme can significantly accelerate the convergence of the iterative algorithms in spatially non-stationary ELAA channels.
Moreover, the computational complexity of the UW-SVD is comparatively minor in relation to the inherent complexity of the iterative algorithms.
\end{abstract}                                                                                                                       
\begin{IEEEkeywords}
User-wise singular value decomposition (UW-SVD), extremely large aperture array (ELAA), multiple-input multiple-output (MIMO), iterative algorithms.
\end{IEEEkeywords}

\section{Introduction}\label{sec1}
In massive multiple-input multiple-output (mMIMO) systems, hundreds of service antennas are deployed to serve tens of user antennas simultaneously in the same frequency band \cite{Ngo2014}.
The optimal MIMO detector, called maximum likelihood sequence detection, faces an excessively high complexity in massive MIMO systems due to the exponential growth in complexity \cite{Albreem2019}.
Linear MIMO detectors such as zero-forcing (ZF) and regularized ZF (RZF) present more manageable complexity levels.
However, they are still prohibitive for practical use due to the necessity of channel matrix inversion.
A number of iterative algorithms have been proposed to offer linear detector performance bypassing the matrix inverse \cite{Yang2015}.                                                                                                                                                  
These algorithms primarily belong to four categories: stationary iterative (SI) methods \cite{Zhang2021}, gradient methods \cite{Yin2014}, quasi-Newton (QN) methods \cite{Li2022a}, and belief propagation \cite{Lyu2019}. 
They demonstrate fast convergence when dealing with well-conditioned channel matrices.

In the emerging MIMO paradigm deployed with an extremely large aperture arrays (ELAA), the user terminals (UTs) are located in the near field of the service antenna array \cite{Bjoernson2017}.
ELAA channels can be very ill-conditioned due spatial non-stationarity the presence of line-of-sight (LoS) links.
Additionally,  each UT in the ELAA-MIMO system is typically equipped with multiple antennas, resulting in high intra-user correlations and further exacerbating the ill-conditioned nature of the ELAA channels.
Consequently, iterative algorithms that exhibit rapid convergence in conventional mMIMO channels demonstrate slower convergence in ELAA channels.
Furthermore, some algorithms, such as Jacobi iteration, fail in ELAA-MIMO systems due to the ill-conditioned nature of channels.
This highlights the need for developing efficient iterative MIMO detectors for future ELAA-MIMO signal detection.

In this paper, we propose a novel iterative detection scheme for ELAA-MIMO systems based on user-wise singular value decomposition (SVD).
By conducting SVD on each UT's sub-channel, the MIMO signal model can be transformed into an equivalent form with a better-conditioned transfer function. 
Subsequently, existing iterative algorithms can be applied to recover the transmitted signal from the modified signal model toward ZF performance.
It is demonstrated that, through computer simulations, the proposed UW-SVD scheme can significantly accelerate the convergence rate of existing iterative algorithms, such as Jacobi iteration (JI), Gauss-Seidel (GS), symmetric successive over-relaxation (SSOR), and limited memory Broyden–Fletcher–Goldfarb–Shanno (L-BFGS).
Moreover, the computational complexity of the UW-SVD is comparatively minor in relation to the inherent complexity of the iterative algorithms.

\section{Signal Model, Preliminaries and Problem Formulation}\label{sec2}
\subsection{Signal Model}
Let $M$ and $N$ denote the number of service antennas and user terminal (UT) antennas, respectively.
The uplink signal model for ELAA-MIMO and conventional MIMO systems can both be expressed as follows
\begin{equation}\label{eqn01}
	\mathbf{y} =\mathbf{H} \mathbf{s} + \mathbf{z},
\end{equation}
where $\mathbf{y} \in \mathbb{C}^{M \times 1}$ denotes the received signal vector, $\mathbf{s} \in \mathbb{C}^{N \times 1}$ the transmitted signal vector, $\mathbf{z} \sim \mathcal{CN}(0,\sigma_z^2\mathbf{I}_M)$ the \gls{awgn}, and $\mathbf{I}_M$ is an $(M)\times(M)$ identity matrix.
In conventional mMIMO systems, every element in $\mathbf{H}\in \mathbb{C}^{M \times N}$ is assumed to follow and independent and identically distributed (i.i.d.) Rayleigh fading. 
However, in ELAA-MIMO systems, the wireless channels become spatially non-stationary and can consist of a mixture of LoS/Non-LoS (NLoS) links \cite{Liu2021}. 
More specifically, the NLoS links should be extended to obey i.n.d. Rayleigh fading as follows \cite{Amiri2018}
\begin{equation}\label{eqn02}
	h_{m,n}^{\textsc{nlos}} \triangleq \left(\frac{\beta_{0}}{d_{m,n}^{\gamma_{0}}}\right) \omega_{m,n};
\end{equation}
and in the LoS links should be extended to obey i.n.d. Rice fading as follows \cite{Wang2022}
\begin{equation}
	h_{m,n}^{\textsc{los}} \triangleq \dfrac{\beta_{1}}{d_{m,n}^{\gamma_{1}}}\left(\sqrt{\dfrac{\kappa}{\kappa + 1}} \phi_{m,n} + \sqrt{\dfrac{1}{\kappa + 1}}\omega_{m,n}\right).
\end{equation}
where $\beta_{0}$ and $\beta_{1}$ denote the path-loss coefficient in NLoS and LoS state, respectively; $\gamma_{0}$ and $\gamma_{1}$ the path-loss exponent in NLoS and LoS state, respectively; $d_{m,n}$ the distance between the $m$-th transmit antenna and the $n$-th receive antenna; 
$\phi_{m,n}$ the phase of LoS path; and $\kappa \sim \mathcal{LN}(\mu_{\kappa}, \sigma_{\kappa}^{2})$ indicates the Rice K-factor \cite{3gpp.38.901}. The random distribution of LoS/NLoS state is described by exponentially decaying windows in \cite{Liu2021}.

\subsection{Preliminaries}
Let $\widehat{\mathbf{s}}$ represent the estimation vector of ZF detector, it can be expressed as follows \cite{Albreem2019}
\begin{equation}\label{eqn03}
	\widehat{\mathbf{s}} = (\mathbf{H}^H\mathbf{H})^{-1}\mathbf{H}^H\mathbf{y} = \mathbf{H}^{\dagger} \mathbf{y},
\end{equation}
where ${(\cdot)}^{\dagger}$ denotes the pseudo-inverse of the input matrix.
To avoid the high complexity matrix inverse, the problem in \eqref{eqn03} can be rewritten as the following linear form
\begin{equation}\label{eqn04}
	\mathbf{\bar{A}} \widehat{\mathbf{s}} = \mathbf{\bar{b}},
\end{equation}
where $\mathbf{\bar{A}} \triangleq \mathbf{H}^{H}\mathbf{H}$ and $\mathbf{\bar{b}} \triangleq \mathbf{H}^{H}\mathbf{y}$.
Numerous iterative algorithms have been proposed to find $\widehat{\mathbf{s}}$ with a square-order complexity \cite{Yang2015}. 
They can be expressed as follows
\begin{equation}\label{eqn11322804}
\mathbf{s}_{t+1} = f(\mathbf{s}_{t}; \mathbf{\bar{A}}, \mathbf{\bar{b}}),
\end{equation}
where $f(\cdot)$ is a linear function.
Since $\mathbf{\bar{A}}$ is a symmetric positive-definite (SPD) matrix, the convergence of $f(\cdot)$ for gradient and QN methods can be guaranteed \cite{Yang2015}, which means
\begin{equation} \label{eqn22472504}
	\lim\limits_{t \rightarrow \infty} \mathbf{s}_{t} \rightarrow \widehat{\mathbf{s}}.
\end{equation}
For SI methods, the convergence can only be guaranteed when the spectral radius of their iterative matrices is less than $1$ \cite{Wang2022a}.
However, they will all suffer from slow convergence in the ill-conditioned ELAA channels.

\subsection{Problem Formulation}
As previously mentioned, existing iterative algorithms can only offer fast convergence in well-conditioned MIMO channels.
In the ELAA-MIMO systems, their convergence becomes slow or even fail due to the channel ill-conditioning.
This necessitates the development of fast iterative MIMO detectors for future ELAA-MIMO signal detection.

\section{UW-SVD Assisted Iterative ELAA-MIMO Detections}
In this section, we will discuss the concept of UW-SVD and and its application in accelerating existing iterative MIMO detection techniques. 
Additionally, some minor modification of the existing iterative algorithms will be discussed.

\subsection{The Concept of UW-SVD}
Denote $K$ as the number of UT, suppose that each UT is deployed with $N_k$ transmit antennas, we have $\sum_{k=1}^{K} N_k = N$.
The channel matrix can be written as the concatenated way as $\mathbf{H} = [\mathbf{H}_{1}, ... \mathbf{H}_{K}]$, where $\mathbf{H}_{k} \in \mathbb{C}^{M\times N_k}$ denotes the sub-channel of the $k^{th}$ UT.
The main idea of UW-SVD is performing economy-size SVD on $\mathbf{H}_{k}, \forall k$ as follows
\begin{equation}\label{eqn13372804}
\mathbf{H}\ =  [\mathbf{U}_{1}\mat{\Sigma}_{1}\mathbf{V}_{1}^{H}, ... , \mathbf{U}_{K}\mat{\Sigma}_{K}\mathbf{V}_{K}^{H}],
\end{equation}
where $\mathbf{U}_k \in \mathbb{C}^{M \times N_k}$ is a unitary matrix, $\mathbf{\Sigma}_k \in \mathbb{C}^{M \times N_k}$ the diagonal matrix containing the singular values of $\mathbf{H}_{k}$, and $\mathbf{V}_k \in \mathbb{C}^{N_k \times N_k}$ is a unitary matrix.
According to \eqref{eqn13372804}, $\mathbf{H}$ can be rewritten as the product of the following three matrices
\begin{equation}
\mathbf{H} = \mathbf{\Psi} \mathbf{\Sigma} \mathbf{V}^{H}
\end{equation}
where $\mathbf{\Psi} \triangleq [\mathbf{U}_{1}, ..., \mathbf{U}_{K}]$, $\mathbf{\Sigma} \triangleq \mathrm{diag}[\mathbf{\Sigma}_{1}, ... , 	\mathbf{\Sigma}_{K}]$, $\mathbf{V} \triangleq \mathrm{diag}[\mathbf{V}_{1}, ... , 	\mathbf{V}_{K}]$, and $\mathrm{diag}(\cdot)$ concatenates the input matrices in a block diagonal manner.

\begin{rmk}
	Note that the matrix $\mathbf{\Psi}$ is not an unitary matrix, and the signal model in \eqref{eqn01} can be rewritten as follows
	\begin{equation} \label{eqn16542104}
		\mathbf{y} = \mathbf{\Psi} \mathbf{x} + \mathbf{z}, 
	\end{equation}
	where $\mathbf{x} \triangleq \mathbf{\Sigma}\mathbf{V}^{H} \mathbf{s}$.
 	With this equivalent signal model, we can recover $\mathbf{x}$ from $\mathbf{y}$ and then recover $\mathbf{s}$ from $\mathbf{x}$.
\end{rmk}
	
\subsection{The Proposed UW-SVD Scheme}
\begin{figure}[t]
	\centering
	\includegraphics[width=0.46\textwidth]{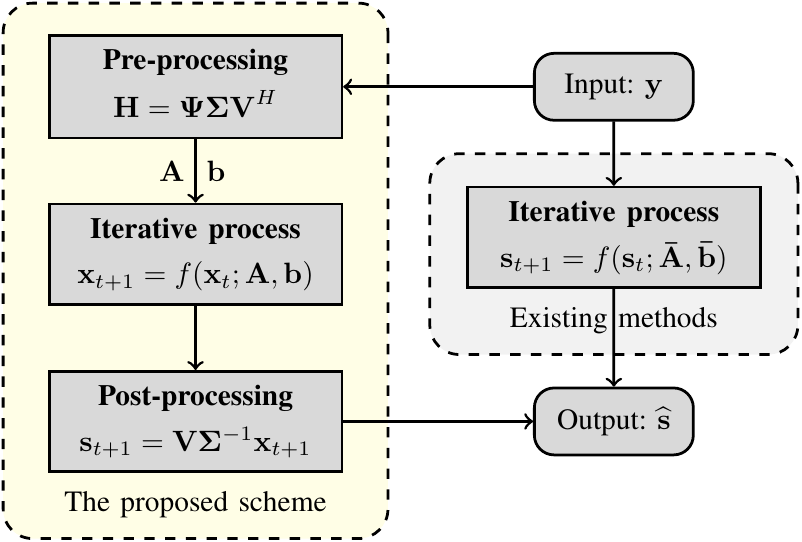}
	\caption{A comparison between the proposed UW-SVD scheme and current iterative algorithms. The UW-SVD scheme incorporates both a pre-processing step and a post-processing step.}
	\label{fig01}
	\vspace{-1em}
\end{figure}

\begin{thm} \label{thm01}
Given $\mathbf{H} = \mathbf{\Psi}\mathbf{\Sigma}\mathbf{V}^{H}$, the estimation of $\mathbf{x}$ using the ZF principle can be expressed as
\begin{equation} \label{eqn16242304}
\widehat{\mathbf{x}} = \mathbf{\Psi}^{\dagger} \mathbf{y},
\end{equation}
Subsequently, $\mathbf{s}$ can be recovered from $\mathbf{x}$ as follows
\begin{equation} \label{eqn10252904}
\widetilde{\mathbf{s}} = \mathbf{V}\mathbf{\Sigma}^{-1} \widehat{\mathbf{x}}.
\end{equation}
The estimation of $\mathbf{s}$ based on the UW-SVD equivalent signal model is the same at that of ZF detector as follows
\begin{equation} \label{eqn10412904}
\widetilde{\mathbf{s}} = \widehat{\mathbf{s}}.
\end{equation}
\end{thm}

\begin{IEEEproof}
	See \textsc{Appendix} \ref{app01}
\end{IEEEproof}

For notational simplicity, $\widehat{\mathbf{s}}$ will be used to represent the estimation vector in the UW-SVD assisted algorithms.

\begin{cor} \label{cor1.1}
Define $\mathbf{A} \triangleq \mathbf{\Psi}^{H}\mathbf{\Psi}$ and $\mathbf{b} \triangleq \mathbf{\Psi}^{H}\mathbf{y}$, according to \eqref{eqn16242304}, we can obtain the following linear equation 
\begin{equation}
\mathbf{A} \widehat{\mathbf{x}} = \mathbf{b}.
\end{equation}
Therefore, low-complexity iterative algorithms can be used to recover $\widehat{\mathbf{x}}$ bypassing the $\mathbf{A}$ inverse as follows
\begin{equation}
	\mathbf{x}_{t+1} = f(\mathbf{x}_{t}; \mathbf{A}, \mathbf{b}).
\end{equation}
For any $f(\cdot)$ converging to $\widehat{\mathbf{x}}$, the following holds
\begin{equation}
\lim\limits_{t \rightarrow \infty} \mathbf{V}\mathbf{\Sigma}^{-1}\mathbf{x}_{t+1} = \widehat{\mathbf{s}}.
\end{equation} 
\end{cor}

\begin{IEEEproof}
The proof is straightforward and omitted here.
\end{IEEEproof}

According to \textit{Corollary} \ref{cor1.1}, we propose a UW-SVD scheme that can be used to accelerate the convergence of existing iterative algorithms.
\figref{fig01} shows the block diagram of the proposed UW-SVD scheme and conventional iterative algorithms.
Compared to conventional iterative MIMO detectors, UW-SVD assisted iterative algorithm have two more steps: pre-processing and post-processing.
In the iterative process, $f(\cdot)$ can be replaced by SI, gradient or SN methods.
Next, we will present some choices of existing iterative algorithms in the proposed UW-SVD scheme.

\subsection{The Choices of $f(\cdot)$ in the Iterative Process}
A large number of iterative algorithms can be used as the choice of $f(\cdot)$ in the proposed UW-SVD scheme.
In this section, we will present some representative examples, which can be further simplified in our proposed scheme, i.e, SI methods, and the L-BFGS algorithm.
For the simplification of notation, we define
\begin{equation}
	\mathbf{g}_{t} \triangleq \mathbf{A}\mathbf{x}_{t} - \mathbf{b},
\end{equation}
which is called residual vector in SI methods and gradient vector in gradient and QN methods.

SI methods, including Richardson iteration (RI), JI, GS, and SSOR, are based on the following matrix splitting
\begin{equation}
	\mathbf{A} = \mathbf{D} + \mathbf{L} + \mathbf{L}^{H},
\end{equation}
where $\mathbf{D}$ and $\mathbf{L}$ denote the diagonal and strict lower triangular parts of $\mathbf{A}$, respectively.
They can all be expressed as follows
\begin{equation}
	\mathbf{x}_{t+1} = \mathbf{x}_{t} - \mathbf{M}^{-1}\mathbf{g}_{t}.
\end{equation} 
where $\mathbf{x}_{t}$ denotes the $t^{th}$ estimation vector, and $\mathbf{M}$ is the preconditioning matrix to distinguish between different SI methods: $\mathbf{M}_{\textsc{ri}} = \mathbf{I}_{N}$, $\mathbf{M}_{\textsc{ji}} = \mathbf{D}$, $\mathbf{M}_{\textsc{gs}} = \mathbf{D} + \mathbf{L}$ \cite{Zhang2021}, and $\mathbf{M}_{\textsc{ssor}} = \left(\mathbf{D} + \mathbf{L}\right)\mathbf{D}^{-1}\left(\mathbf{D} + \mathbf{L}\right)^H$ \cite{Xie2016}.
According to the special structure of $\mathbf{\Psi}$, the diagonal of $\mathbf{A}$ is an identity matrix, i.e., $\mathbf{D} = \mathbf{I}_{N}$.
Therefore, in the UW-SVD scheme RI is equivalent to JI and $\mathbf{M}_{\textsc{ssor}} = \left(\mathbf{I}_{N} + \mathbf{L}\right)\left(\mathbf{I}_{N} + \mathbf{L}\right)^H$.

L-BFGS algorithm is one of the most effective QN methods in MIMO detection, which can be expressed as follows
\begin{equation}\label{eqn07}
	\mathbf{x}_{t+1} = \mathbf{x}_{t} - \xi_{t} \mathbf{d}_{t},
\end{equation}
where $\mathbf{d}_{t}$ denotes the update direction, and it is given by \cite{Li2022}
\begin{equation} \label{eqn09322704}
	\mathbf{d}_{t} = -\mat{\Theta}\mathbf{g}_{t} + \dfrac{(\mathbf{x}_{t} - \mathbf{x}_{t-1})(\mathbf{g}_{t} - \mathbf{g}_{t-1})^{H}\mat{\Theta}\mathbf{g}_{t}}{(\mathbf{x}_{t} - \mathbf{x}_{t-1})^{H}(\mathbf{g}_{t} - \mathbf{g}_{t-1})},
\end{equation}
where $\mat{\Theta}$ is the initial Hessian matrix.
It is usually set as the diagonal part of $\mathbf{A}$.
In the proposed UW-SVD scheme, the diagonal elements of $\mathbf{A}$ is an identity matrix.
Therefore, the update direction of L-BFGS method can be further simplified by removing $\mat{\Theta}$ from \eqref{eqn09322704}.
Moreover, the step size $\xi_{t}$ can be expressed as follows \cite{Li2022}
\begin{equation}
\xi_{t} = \dfrac{\mathbf{g}_{t}^{H}\mathbf{d}_{t}}{\mathbf{d}_{t}^{H}\mathbf{A}\mathbf{d}_{t}}.
\end{equation}

In this section, we present a few representative examples of existing iterative MIMO detectors that can be further simplified using the proposed UW-SVD scheme.
Due to the page limitation, we are unable to cover all the compatible iterative algorithms, such as gradient descent, conjugate gradient, and damped JI.
Next we will focus on the complexity and performance analysis of the proposed scheme.

\subsection{Complexity Analysis of UW-SVD Scheme}
The complexity of conducting SVD for $\mathbf{H}_{k}$ is $MN_{k}^{2}$.
For the simplification of complexity analysis, we assume that every UT has the same antennas number, i.e., $N_{k} = N_{\textsc{ut}}, \forall k$.
Therefore, the total complexity of the pre-processing in UW-SVD scheme is $KMN_{\textsc{ut}}^{2}$ or $N_{\textsc{ut}}MN$, since we have $N = KN_\textsc{ut}$.
In the post-processing, the complexity of calculating $\mat{\Sigma}^{-1}$ is $N$, since it is a diagonal matrix.
The complexity of calculating $\mathbf{V}[\mat{\Sigma}^{-1}\mathbf{x}_{t+1}]$ is $KN_{\textsc{ue}}^{2}$ or $NN_{\textsc{ue}}$.
Therefore, the total complexity of pre-processing and post-processing of the UW-SVD scheme is $N_{\textsc{ut}}(MN + N)$.
In ELAA-MIMO systems, the number of $N_\textsc{ut}$ is usually very small, e.g., $N_{\textsc{ut}} = 2$ or $4$.
Therefore, the computational cost of UW-SVD scheme is still in the square order, which is scalable as $N$ increases.

\subsection{Performance Analysis} \label{sec3e}
\begin{thm}\label{thm02}
Suppose that every elements of $\mathbf{H}$ follow i.i.d. Rayleigh fading distribution, given $K$ and $N_{k}, \forall k$, the condition number of $\mathbf{A}$ is asymptotically same as that of $\mathbf{\bar{A}}$ when $M$ tends to infinity, i.e.,
\begin{equation}
\lim\limits_{M \rightarrow \infty} \mathrm{cond}(\mathbf{A}) = \mathrm{cond}({\mathbf{\bar{A}}}),
\end{equation}
where $\mathrm{cond}(\cdot)$ denotes the condition number of input matrix.
\end{thm}

\begin{IEEEproof}
See \textsc{Appendix} \ref{app02}.
\end{IEEEproof}

The convergence rate of iterative MIMO detectors are dominated by the condition number of the transfer function \cite{Albreem2019}.
\textit{Theorem} \ref{thm02} implies that the convergence rate of iterative algorithms using $\mathbf{A}$ or $\mathbf{\bar{A}}$ can have the same convergence rate in i.i.d. Rayleigh fading channels.
Therefore, the proposed UW-SVD scheme is not specific for conventional mMIMO channels.
On the contrary, our numerical results show that the condition number of $\mathbf{\bar{A}}$ is much lower than that of $\mathbf{\bar{A}}$ in spatially non-stationary ELAA channels.
Therefore, the proposed UW-SVD scheme can offer significantly faster convergence in ELAA-MIMO systems, compared to existing iterative algorithms.

\section{Simulation and Numerical Results}
In this section, the objective are: \textit{1)} to showcase that the UW-SVD assisted iterative algorithms converge faster than current iterative algorithms in ELAA channels, and \textit{2)} to demonstrate that the condition number of $\mathbf{A}$ and $\mathbf{\bar{A}}$ in both conventional MIMO and ELAA-MIMO channels.
For the ELAA channel, we consider the non-stationary fading channel model proposed in \cite{Liu2021} to conduct the Monte Carlo trials.
In our simulation and numerical results, it is assumed that there are $256$ service antennas configured as a uniformly linear array with equal spacing at half of the wavelength at the center frequency of $3.5$ $\mathrm{GHz}$.
There are $32$ UTs each with $2$ transmitter antennas (i.e., $N = 64$) are deployed parallel to the ELAA with equal spacing of $1$ meter.
The wireless environment is assumed to be urban-micro street canyon.
According to the third Generation Partnership Project (3GPP) standard \cite{3gpp.38.901}, we have $\beta_{0} = 0.020$, $\gamma_{0} = 1.765$, $\beta_{1} = 0.007$, $\gamma_{1} = 1.050$, $\mu_{\kappa} = 9$ $\mathrm{dB}$ and $\sigma_{\kappa} = 10$ $\mathrm{dB}$.
The modulation for the simulation results is set as $16$ $\mathrm{QAM}$.
The objectives set the following two experiments.

{\bf Experiment 1:} 
\begin{figure*}[t]
	\subfigure[Spatially non-stationary ELAA channel \cite{Liu2021}; Es/No = $22$ $\mathrm{dB}$]{
			\label{fig02a}
			\centering
			\hspace{1em}
			\includegraphics[width=0.42\textwidth]{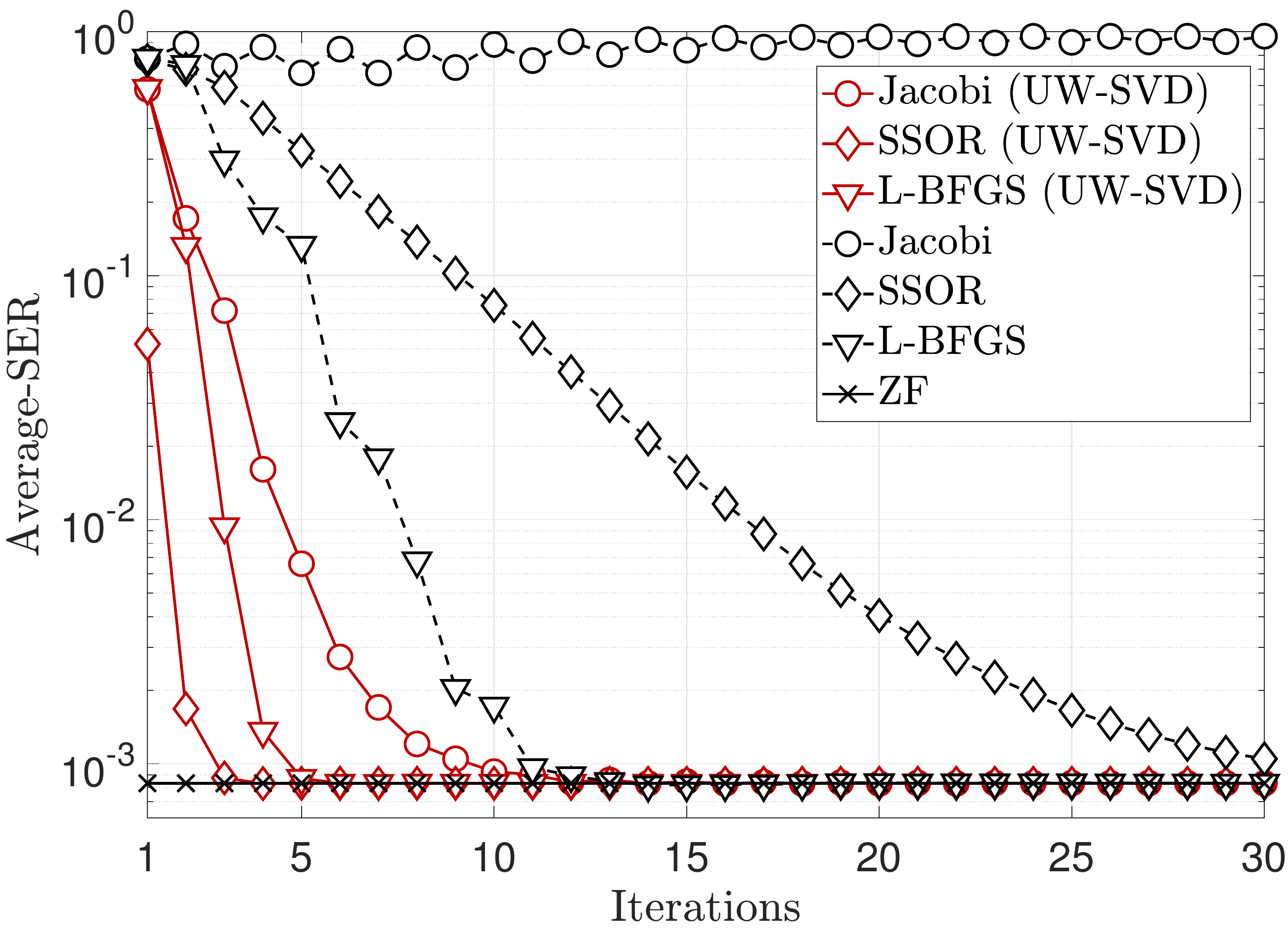}
			\vspace{1em}}
			\hspace{3em}
	\subfigure[i.i.d. Rayleigh fading channel; Es/No = $19$ $\mathrm{dB}$]{
			\label{fig02b}
			\centering
			\includegraphics[width=0.42\textwidth]{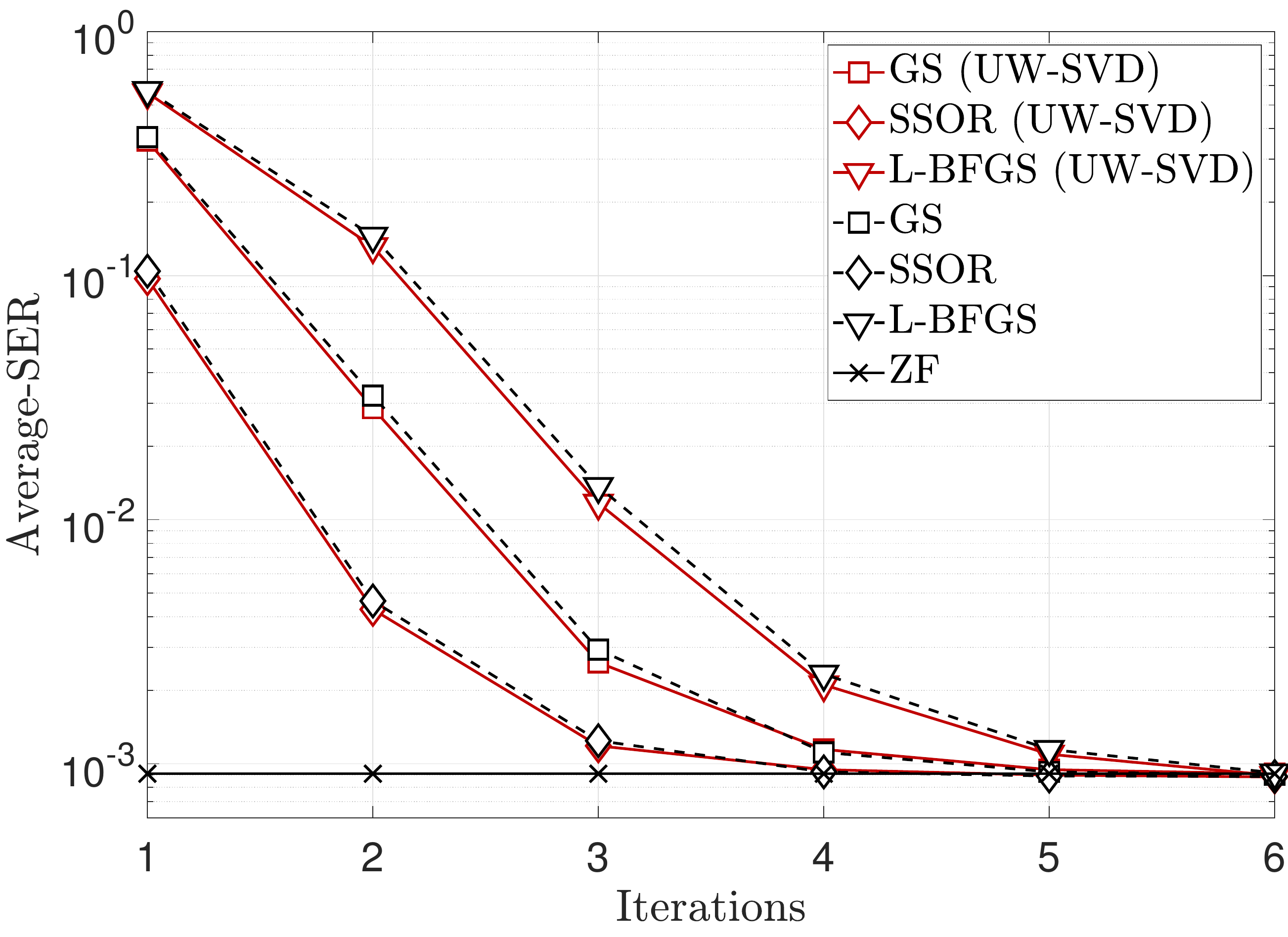}
			\vspace{1em}}		
	\caption{\label{fig02} The SER performance comparison between the proposed UW-SVD assisted iterative algorithms and conventional iterative algorithms in $256 \times 64$ MIMO systems with $16$-QAM modulation.}
	\vspace{-1em}
\end{figure*}
This experiment aims to demonstrate that UW-SVD assisted iterative algorithms convergence faster than the existing iterative algorithms in spatially non-stationary ELAA channels.
\figref{fig02} shows the average symbol error rate (SER) versus the iterations of various iterative MIMO detectors in both conventional MIMO and ELAA-MIMO channels.
As shown in \figref{fig02a}, it is obviously that the iterative algorithms assisted by the proposed UW-SVD scheme converge significantly faster than the original iterative algorithms in ELAA channels.
For instances, the original SSOR algorithm requires $30$ iterations to achieve ZF detection performance, whereas with the assistance of the proposed UW-SVD scheme, it can reach ZF detection performance in just $4$ iterations; the L-BFGS method converges twice as fast when leveraging the proposed UW-SVD scheme.
Moreover, it can be found that the UW-SVD assisted JI converge faster than L-BFGS, while the original JI fails due to the channel ill-conditioning.

\figref{fig02b} shows the convergence of several iterative algorithms in conventional MIMO systems with i.i.d. Rayleigh fading channel.
It can be observed that the UW-SVD scheme can only offer slightly faster convergence, compared to the original algorithms.
This is inline with our theoretical analysis in \secref{sec3e}.
Moreover, if we compare \figref{fig02a} and \figref{fig02b}, it can be found that the convergence of UW-SVD assisted iterative algorithms in ELAA channels are similar to the original algorithms in i.i.d. Rayleigh fading channels.
For instances, UW-SVD assisted SSOR requires the $4$ iterations in ELAA channels and the original SSOR also requires $4$ iterations in conventional MIMO channels.
These results indicate that, by leveraging the proposed UW-SVD scheme, linear iterative algorithms can achieve convergence rates as fast as they do in conventional MIMO channels.

{\bf Experiment 2:} 
\begin{figure}[t]
	\centering
	\subfigure{
			\label{fig03b}
			\centering
			\includegraphics[width=0.42\textwidth]{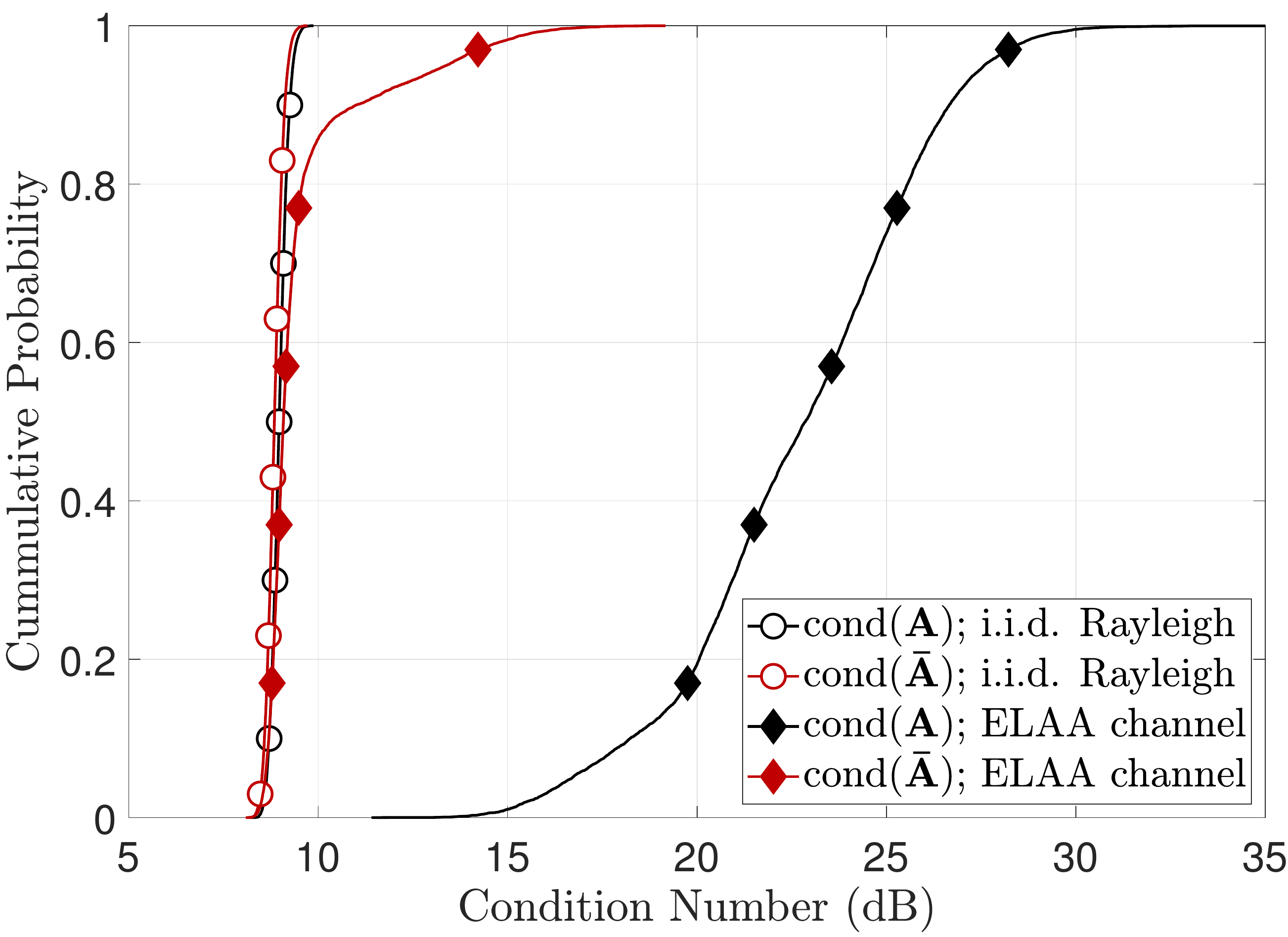}}		
	\caption{\label{fig03} The CDF of the condition numbers for $\mathbf{A}$ and $\mathbf{\bar{A}}$ in conventional MIMO and ELAA-MIMO channels.}
	\vspace{-1em}
\end{figure}
The objective of this experiment is to show the cumulative distribution function (CDF) of $\mathrm{cond}(\mathbf{A})$ and $\mathrm{cond}(\mathbf{\bar{A}})$ in both conventional MIMO and ELAA-MIMO channels.
As depicted in \figref{fig03}, the CDF distributions of $\mathrm{cond}(\mathbf{A})$ and $ \mathrm{cond}(\mathbf{\bar{A}})$ are observed to be very similar to one another in i.i.d. Rayleigh fading channels.
This result validates our theoretical result presented in \textit{Theorem} \ref{thm02}.
This is the underlying reason why UW-SVD assisted algorithms exhibit similar convergence rates when compared to the original iterative algorithms.
On the contrary, in the ELAA-MIMO channels, it can be found that $\mathrm{cond}(\mathbf{A})$ tends to be much larger than $\mathrm{cond}(\mathbf{\bar{A}})$.
Therefore, the proposed UW-SVD scheme can significantly accelerate the convergence of existing iterative algorithm in spatially non-stationary ELAA channels.

\section{Conclusion and Outlook}
In this paper, we propose a novel UW-SVD scheme which can accelerate the convergence of existing iterative MIMO detectors in spatially non-stationary ELAA channels.
The scheme consists of three steps: \textit{1)} in the pre-processing step, the MIMO signal is converted into an equivalent form with a better-conditioned transfer function; \textit{2)} during the iterative process, existing iterative algorithms are utilized to recover $\mathbf{x}$; and \textit{3)} in the post-processing step, the transmitted signal is recovered from $\mathbf{x}$. 
It is demonstrated that UW-SVD assisted iterative algorithms, including JI, GS, SSOR, and L-BFGS methods, converge significantly faster (at least two times) in ELAA channels compared to their original performances.

Due to the page limit, we only present the UW-SVD scheme's ability to achieve ZF detection performance. 
However, this can be extended to more general cases for RZF detectors, such as the linear minimum mean square error (LMMSE) detector. 
Moreover, certain current MIMO detectors, like approximate message passing (AMP), cannot be directly combined with the proposed UW-SVD scheme. 
These interesting results, along with others, will be presented in our transaction version.

\appendices
\section{Prove of \textit{Theorem \ref{thm01}}} \label{app01}
According to \eqref{eqn16242304} and \eqref{eqn10252904}, $\widetilde{\mathbf{s}}$ can be expressed as follows
\begin{equation}
	\widetilde{\mathbf{s}} = \mathbf{V}\mathbf{\Sigma}^{-1} \mathbf{\Psi}^{\dagger}\mathbf{y},
\end{equation}
where $\mathbf{V}\mathbf{\Sigma}^{-1} \mathbf{\Psi}^{\dagger}$ can be expressed as follows
\begin{IEEEeqnarray} {ll}
	\mathbf{V}\mathbf{\Sigma}^{-1}\mathbf{\Psi}^{\dagger}\	&=\mathbf{V}\mathbf{\Sigma}^{-1}(\mathbf{\Psi}^{H}\mathbf{\Psi})^{-1}\mathbf{\Psi}^{H}, \nonumber \\
	&=\mathbf{V}\mathbf{\Sigma}^{-1}(\mathbf{\Psi}^{H}\mathbf{\Psi})^{-1}\mathbf{\Sigma}^{-1}\mathbf{V}^{H} \mathbf{V}\mathbf{\Sigma}\mathbf{\Psi}^{H}, \nonumber \\	
	&=\big[(\mathbf{\Psi}\mathbf{\Sigma}\mathbf{V}^{H})^{H}(\mathbf{\Psi}\mathbf{\Sigma}\mathbf{V}^{H})\big]^{-1} \mathbf{V}\mathbf{\Sigma}\mathbf{\Psi}^{H}.	\label{eqn10382904}
\end{IEEEeqnarray}
Plugging $\mathbf{H} = \mat{\Psi}\mathbf{\Sigma}\mathbf{V}^{H}$ into \eqref{eqn10382904} yields
\begin{equation}
	\mathbf{V}\mathbf{\Sigma}^{-1}\mathbf{\Psi}^{\dagger} = (\mathbf{H}^{H}\mathbf{H})^{-1}\mathbf{H}^{H} = \mathbf{H}^{\dagger}.
\end{equation}
According to \eqref{eqn03}, \eqref{eqn10412904} can be obtained and \textit{Theorem} \ref{thm01} is proved.

\section{Prove of \textit{Theorem \ref{thm02}}} \label{app02}
At first, let us focus on $\mathbf{H}_{k} = \mathbf{U}_{k}\mathbf{\Sigma}_{k}\mathbf{V}_{k}^{H}$.
According to the favorable propagation \cite{Chen2018}, given the number of antennas per UT, the columns of the intra-user channel vectors are asymptotically orthogonal as follows
\begin{equation}
\lim\limits_{M \rightarrow \infty} \mathbf{\Sigma}_{k} = \sigma^{2}_{h} \mathbf{I}_{N_k},
\end{equation}
where $\sigma^{2}_{h}$ denotes the variance of each channel elements.
Therefore, we have
\begin{equation}
\lim\limits_{M \rightarrow \infty} \mathbf{H}_{k} = \sigma^{2}_{h} \mathbf{U}_{k} \mathbf{V}_{k}^{H},
\end{equation}
The whole channel matrix can be represented as follows
\begin{equation} \label{eqn11552904}
\lim\limits_{M \rightarrow \infty} \mathbf{H} = \sigma_{h}^{2} \mathbf{\Psi} \mathbf{V}^{H}.
\end{equation}
The right term $\mathbf{V}$ in \eqref{eqn11552904} is an unitary matrix. 
According to the property of unitary matrix, it is clear that the condition number of $\mathbf{\Psi}$ is the same as that of $\mathbf{H}$ and \textit{Theorem} \ref{thm02} is proved.

\section*{Acknowledgement}
This work is partially funded by the 5G Innovation Centre and 6G Innovation Centre.

\ifCLASSOPTIONcaptionsoff
\newpage
\fi

\bibliographystyle{IEEEtran}
\bibliography{IEEEabrv,mMIMO} 
\begin{thebibliography}{10}
\providecommand{\url}[1]{#1}
\csname url@samestyle\endcsname
\providecommand{\newblock}{\relax}
\providecommand{\bibinfo}[2]{#2}
\providecommand{\BIBentrySTDinterwordspacing}{\spaceskip=0pt\relax}
\providecommand{\BIBentryALTinterwordstretchfactor}{4}
\providecommand{\BIBentryALTinterwordspacing}{\spaceskip=\fontdimen2\font plus
\BIBentryALTinterwordstretchfactor\fontdimen3\font minus
  \fontdimen4\font\relax}
\providecommand{\BIBforeignlanguage}[2]{{%
\expandafter\ifx\csname l@#1\endcsname\relax
\typeout{** WARNING: IEEEtran.bst: No hyphenation pattern has been}%
\typeout{** loaded for the language `#1'. Using the pattern for}%
\typeout{** the default language instead.}%
\else
\language=\csname l@#1\endcsname
\fi
#2}}
\providecommand{\BIBdecl}{\relax}
\BIBdecl

\bibitem{Ngo2014}
H.~Q. Ngo, E.~G. Larsson, and T.~L. Marzetta, ``Aspects of favorable
  propagation in massive {MIMO},'' in \emph{22nd European Signal Processing
  Conference (EUSIPCO)}, 2014, pp. 76--80.

\bibitem{Albreem2019}
M.~A. Albreem, M.~Juntti, and S.~Shahabuddin, ``Massive {MIMO} detection
  techniques: A survey,'' \emph{{IEEE} Commun. Surveys Tuts.}, vol.~21, no.~4,
  pp. 3109--3132, 4th Quart. 2019.

\bibitem{Yang2015}
S.~Yang and L.~Hanzo, ``Fifty years of {MIMO} detection: The road to
  large-scale {MIMOs},'' \emph{{IEEE} Commun. Surveys Tuts.}, vol.~1, no.~4,
  pp. 1941--1988, 4th Quart. 2015.

\bibitem{Zhang2021}
C.~Zhang, Z.~Wu, C.~Studer, Z.~Zhang, and X.~You, ``Efficient soft-output
  {Gauss–Seidel} data detector for massive {MIMO} systems,'' \emph{{IEEE}
  Trans. Circuits Syst. {I}}, vol.~68, no.~12, pp. 5049--5060, Dec. 2021.

\bibitem{Yin2014}
B.~Yin, M.~Wu, J.~R. Cavallaro, and C.~Studer, ``Conjugate gradient-based
  soft-output detection and precoding in massive {MIMO} systems,'' in
  \emph{Proc. IEEE Global Commun. Conf. (GLOBECOM)}, 2014, pp. 3696--3701.

\bibitem{Li2022a}
{L. Li} and J.~Hu, ``Fast-converging and low-complexity linear massive {MIMO}
  detection with {L-BFGS} method,'' \emph{{IEEE} Trans. Veh. Technol.},
  vol.~71, no.~10, pp. 10\,656--10\,665, Oct. 2022.

\bibitem{Lyu2019}
S.~Lyu and C.~Ling, ``Hybrid vector perturbation precoding: The blessing of
  approximate message passing,'' \emph{{IEEE} Trans. Signal Process.}, vol.~67,
  no.~1, pp. 178--193, Oct. 2019.

\bibitem{Bjoernson2017}
E.~{Bj\"ornson}, J.~{Hoydis}, and L.~{Sanguinetti}, ``Massive {MIMO} networks:
  Spectral, energy, and hardware efficiency,'' \emph{Foundations and
  Trends{\textregistered} in Signal Processing}, vol. Nov., no. 3-4, pp.
  154--655, 2017.

\bibitem{Liu2021}
J.~Liu, Y.~Ma, J.~Wang, N.~Yi, R.~Tafazolli, S.~Xue, and F.~Wang, ``A
  non-stationary channel model with correlated {NLoS/LoS} states for
  {ELAA-mMIMO},'' in \emph{Proc. IEEE Global Commun. Conf. (GLOBECOM)}, 2021,
  pp. 1--6.

\bibitem{Amiri2018}
A.~{Amiri}, M.~{Angjelichinoski}, E.~{De Carvalho}, and R.~W. {Heath},
  ``Extremely large aperture massive {MIMO}: Low complexity receiver
  architectures,'' in \emph{Proc. IEEE GLOBECOM Workshops (GC Wkshps)}, 2018,
  pp. 1--6.

\bibitem{Wang2022}
J.~Wang, Y.~Ma, N.~Yi, R.~Tafazolli, and F.~Wang, ``Network-{ELAA} beamforming
  and coverage analysis for {eMBB/URLLC} in spatially non-stationary {Rician}
  channels,'' in \emph{Proc. IEEE Int. Conf. Commun. (ICC)}, 2022, pp. 1--6.

\bibitem{3gpp.38.901}
3GPP, ``{Study on channel model for frequencies from 0.5 to 100 GHz},'' {3rd
  Generation Partnership Project (3GPP)}, Tech. Specification 38.901, Dec.
  2019, version 16.1.0.

\bibitem{Wang2022a}
Z.~Wang, R.~M. Gower, Y.~Xia, L.~He, and Y.~Huang, ``Randomized iterative
  methods for low-complexity large-scale {MIMO} detection,'' \emph{{IEEE}
  Trans. Signal Process.}, vol.~70, pp. 2934--2949, Jun. 2022.

\bibitem{Xie2016}
T.~Xie, L.~Dai, X.~Gao, X.~Dai, and Y.~Zhao, ``Low-complexity {SSOR}-based
  precoding for massive {MIMO} systems,'' \emph{{IEEE} Commun. Lett.}, vol.~20,
  no.~4, pp. 744--747, Apr. 2016.

\bibitem{Li2022}
L.~Li and J.~Hu, ``An efficient linear detection scheme based on {L-BFGS}
  method for massive {MIMO} systems,'' \emph{{IEEE} Commun. Lett.}, vol.~26,
  no.~1, pp. 138--142, Oct. 2022.

\bibitem{Chen2018}
Z.~Chen and E.~Björnson, ``Channel hardening and favorable propagation in
  cell-free massive {MIMO} with stochastic geometry,'' \emph{{IEEE} Trans.
  Commun.}, vol.~66, no.~11, pp. 5205--5219, Nov. 2018.

\end{thebibliography}
\end{document}